\documentclass[12pt]{article}
\usepackage[pctex32]{graphics}
\begin{document}
\vspace{1cm}
\begin{center}
~\\
{\bf  \Large Chiral Dynamics and Meson with Non-commutative Dipole Field  in Gauge/Gravity Dual}
\vspace{1cm}

                      Wung-Hong Huang\\
                       Department of Physics\\
                       National Cheng Kung University\\
                       Tainan, Taiwan\\

\end{center}
\vspace{1cm}
\begin{center}{\bf  \Large ABSTRACT } \end{center}
Apply the T-duality and smeared twist to  the D3-brane solution one can  construct the supergravity backgrounds which may dual to supersymmetric or non-supersymmetric non-commutative dipole field theory.  We introduce  D7-brane probe  into the dual supergravity background to study the chiral dynamics and meson spectrum therein.  We first find  that the non-commutative dipole field does not induce the  chiral symmetry breaking even if the supersymmetry was completely broken, contrast to the conventional believing that the chiral symmetry will be broken in the non-supersymmetric theory.  Next, we find that the dipole field does not modify the meson spectrum in the supersymmetric theory while it will reduce the meson bound-state energy in the non-supersymmetric theory.  We also evaluate the static quark anti-quark potential and see that the dipole field has an effect to produce attractive force between the quark and anti-quark.
\vspace{1cm}
\\
\\
\\
\begin{flushleft}
E-mail:  whhwung@mail.ncku.edu.tw\\
\end{flushleft}


\newpage
\section{Introduction}
The original AdS/CFT correspondence [1] only involves fields in the adjoint representation of the gauge group.  To search a supergravity model dual to a realistic gauge dynamics one has to incorporate the quark degrees of freedom. The simplest thing is to add a new type of brane into the configuration in addition to the D3 branes. In six years ago Karch and Katz had proposed to elevate this brane configuration into a supergravity background by introducing  D7 brane probe into the $AdS_5\times S^5$ background [2].  As the string with both ends on the probe brane is dual to quark-antiquark operators the embedding of the brane in $AdS_5\times S^5$ therefore encodes the mass and quark bilinear condensate in the theory. The linearized fluctuations are then dual to mesonic excitations in the gauge theory.   From this point of view the chiral symmetry breaking [3] and meson spectrum [4] have been investigated by studying the behavior of the D7 probe brane on the $AdS_5\times S^5$ background.

Since the supersymmetry forbids the quark condensate we therefore need to use the non-supersymmetric background to study the chiral symmetry breaking.  The first study of chiral dynamics in  [3] used the Constable-Myers geometry [5]. 
Several  scenarios in which supersymmetry breaking leads to chiral symmetry breaking have also been found - for example a set-up of D4 wrapped a circle which describes gauge theory in IR [6], placing the gauge theory on an anti-de-Sitter space [7], on the Lunin-Maldacena background [8,9], or introducing a background magnetic field [10,11].  The meson spectroscopy therein had also been studied (A recent review could be found in [12]) .   

In this paper we will investigate the problem of chiral symmetry breaking and the meson spectroscopy in the non-commutative dipole field form the  gauge/gravity dual.    The proper dual supergravity background had been found in [13,14], in which a nonzero B field shall be with one leg along the brane worldvolume and other transverse to it.  

As the metric adopted to study the meson will be different from that in studying the Wilson loop or giant graviton [15] we will in next section briefly describe the method which enables us to derive the dual supergravity background in the proper coordinate.  In section III we will use the  dual geometry to study the chiral dynamics and meson spectrum.  We find that the dipole field does not induce the chiral symmetry breaking in the supersymmetric and non-supersymmetric theory.   The general believing that the breaking supersymmetry in gravity duals will leads to chiral symmetry breaking [16], however, does not shows in our investigations.  We also find that the dipole field does not modify the meson spectrum in the supersymmetric theory.  Note that the meson spectrum in the non-commutative field theory with Moyal product  had been studied in [17].   They had shown that, despite being a supersymmetric theory, the meson spectrum will be modified by the space non-commutativity.   We have also seen that the dipole field will reduce the meson energy  in the non-supersymmetric theory. In section IV we evaluate the potential energy for a static quark-antiquark pair and see that the dipole field has an effect to produce attractive force between the quark and anti-quark. The last section is devoted to a conclusion.

In the non-commutative dipole field theory each field $\Phi_a$ is associated with a constant dipole length $\ell_a$ and we define the ``non-commutative dipole product" by $\Phi_a (x) * \Phi_b (x) = \Phi_a (x-\ell_b/2) ~\Phi_b (x +\ell_a/2)$ [13].  It is a nonlocal field theory and break Lorentz invariance. Demanding $\Phi_a^\dag * \Phi_a$ to be real will fix the dipole length of $\Phi_a^\dag$ to be minus  of that of $\Phi_a$.  Thus, the gauge field has zero dipole length and the dipole length of an anti-quark will be the negative value of quark, if it is nonzero.  As there is the supergravity solution which dual to the  non-commutative dipole field theory the physical particle may has nonzero  dipole length.  The phenomenal constrain on the value of dipole length, like as that on the noncommutativity of Moyal  product, remains to be found.  Some properties of the non-commutative dipole field theory have been studied in [14,18].  The noncommutative dipole field theories are interesting by themselves and it has a chance of finding a CP violating theory [14]. It is also an appropriate candidate to study the interaction of a neutral particles with finite dipole moments, like neutrinos, with gauge particles like photons. There are some experimental evidences of such interactions, which cannot be described by the commutative version of the standard model of particles [18]. 
\section{Supergravity Solutions}
To find the explicit supergravity background duals to the non-commutative  dipole theory we could start with the following type II supergravity 
solution describing \textsf{N} coincident  D3-brane [19] 
$$ds^2= H(r)^{-1/2}\left[- dt^2+dx^2+dy^2+dz^2\right] + H(r)^{1/2}\left[- \delta_{ab}dw_i^adw_i^b\right],~~~~H(r) = 1+ {\textsf{N}^{4}\over r^4}, \eqno{(2.1)}$$
in which $a,b = 1 \cdot\cdot\cdot 6$.  Following the prescription in [13], we first apply the T-duality transformation on the $z$ axis. Then, consider the ``smeared twist" as we go around the circle of new $z$ axis (with radius $R$), i.e. the ``twisted" compactification will accompany a rotation between $w_1, ..., w_6$ by a matrix $M_{ab}$ in the following way
$$ (t,x,y,z,w_a)\rightarrow (t,x,y,z+ 2\pi R,\sum_{b=1}^6 M_{ab}w_b),~~~~~ a= 1,...6,\eqno{(2.2)}$$
in which $M$ is an element of the Lie algebra $SO(6)$. After the smeared twist we finally  apply the T-duality on the $z$ axis.   The supergravity solution becomes

$$ds^2 = H^{-\frac{1}{2}} \left( -h(r) dt^2 + dx^2 +dy^2 
+ \frac{ d z^2}{ 1 +  \vec w^T M^T M \vec w} \right) +  H^{\frac{1}{2}} \left(d\vec w^T d\vec w - \frac{(d\vec w^T M \vec w )^2}{1 +  \vec w^T M^T M \vec w} \right),\eqno{(2.3)}$$
$$ C^{(4)}_{0123} = H^{-1},~~~~~~e^{2\phi} =  \frac{1}{1 + \vec w^T M^T M \vec w}, ~~~~ \sum_{a=1}^6 B_{za} dw_a = - \frac{ d\vec w^T M \vec w}{1  +  \vec w^T M^T M \vec w}.\eqno{(2.4)}$$
\\
It is known that the theory on the worldvolume of  D3-brane is N = 4 SU(N) SYM theory. The N = 4 SYM theory in four dimensions has 6 real scalars in the representation {\bf 6} of R-symmetry group SU(4) and 4 Weyl fermions in the representation {\bf 4} of SU(4) [1]. To construct the dipole theory we use the R-symmetry charges to determine the dipole vectors of the various fields.  We denote $V^\mu$ as commuting elements of SU(4).  Then, for simplicity, we assume that all dipoles are along $z$ direction and the dipole moment can be determined by $V^3$.  Denoting the matrix representation of  $V^3$ by $U$ for representation {\bf 4} and by $M$ for representation {\bf 6}, the dipole vectors of fermions and scalars are therefore given by the eigenvalues of $U$ and $M$, respectively [13].

  If the eigenvalues of traceless Hermitian $4 \times 4$ matrix $U$ are  $\alpha_1$, $\alpha_2$, $\alpha_3$, $-(\alpha_1+\alpha_2+\alpha_3)$ the associated general form of matrix $M$ can be cast to the following form [13]
$$ M=\pmatrix{0&\alpha_1+\alpha_2&0&0&0&0 \cr -\alpha_1-\alpha_2 
&0&0&0&0&0\cr 0&0&0&\alpha_1+\alpha_3&0&0 \cr 0&0& -\alpha_1
-\alpha_3 &0&0&0\cr 0&0&0&0&0&\alpha_2+\alpha_3 \cr 0&0&0&0& 
-\alpha_2-\alpha_3 &0}.\eqno{(2.5)}$$
This form of matrix $M$ breaks all supersymmetries in general.  However, in the case of  $\alpha_1 = \alpha_2=0$ and $\alpha_3=B$ we can from the eigenvalues of $U$ and $M$  find that 4 bosons have dipole length $\sim 2B$ and 2 fermions have dipole length $\sim B$. These are the content of $N=2$ hyper-multiplet [13].  (Note that the condition  of  zero total dipole length in  Lagrangian will constrain the relative dipole length for each field in the theory [13,14].)  On the other hand, in the case of  $\alpha_1=\alpha_2=\alpha_3= B$ we can from the eigenvalues of $U$ and $M$ find dipole length for each field and see that it describes a non-supersymmetric theory [13].

To proceed, we note that as the metric adopted to study the meson is different from that in studying the Wilson loop or giant graviton [15]  we will use the following coordinates
$$ w_1 = \rho \cos\theta \cos\phi, ~~~~ w_2 = \rho \cos\theta \sin\phi,$$
$$ w_3 = \rho \sin\theta \cos\chi, ~~~~ w_4 = \rho \sin\theta \sin\chi.\eqno{(2.6)}$$
which are difference from those used in [13-15].   Using the above coordinate we can find the background spacetime and investigate the supersymmetric and non-supersymmetric theories, which is presented in the next section.
\section{Chiral dynamics and meson with non-commutative dipole field} 
\subsection{Supersymmetric dipole field theory}
After the evaluations the following supergravity solution is found 
\\
$$ds_{10}^2 = \left(\rho^2 + w_5^2 +w_6^2\right)\left[-dt^2+ dx^2+ dy^2+{ dz^2\over 1+B^2\rho^2}\right]+{1\over \rho^2 + w_5^2 +w_6^2} \left[ dw_5^2 +dw_6^2\right.$$
$$+ d\rho^2 + \rho^2 \left(d\theta^2+ \cos^2\theta d\phi^2 + \sin^2\theta d\chi^2\right)  - {B^2 \left[ \rho^2 \left(\cos^2\theta d\phi + \sin^2\theta d\chi\right)\right]^2\over 1+B^2\rho^2 }. \eqno{(3.1)}$$
$$e^{2\Phi}= {1 \over  1+B^2 \rho^2 },~~~
 \sum_{n_i=\phi,\chi} B_{zi}dn_i = - {B\left[ \rho^2 \left(\cos^2\theta d\phi + \sin^2\theta d\chi\right) \right] \over 1+B^2\rho^2 },\hspace{0cm}\eqno{(3.2)}$$
in the large \textsf{N} limit.  When $B=0$ above result is $AdS_5 \times O^5$ and our background thus describes the $B$ field deformed supergravity spacetime which dual to the non-commutative dipole field theory. Above solution has a nonzero B field with one leg along the brane worldvolume and others transverse to it.  The value $B$ in (3.2) is proportional to the dipole length $\ell$ defined in the ``non-commutative dipole product" 
 $$\Phi_a (x) * \Phi_b (x) = \Phi_a (x-\ell_b/2) ~\Phi_b (x +\ell_a/2). \eqno{(3.3)}$$ for the dipole field $\Phi(x)$ [13].

Following the method of  [2,3] the D7 probe branes are embedded on D3 brane  (with coordinate $(t,x,y,z)$) in such a way that they extend in space-time $(t,x,y,z, \rho, \theta, \phi, \chi)$.  The value  $w_5^2+w_6^2$ specifies the distance between D3 and D7 brane.  The D7 brane configuration described by $w_5(\rho), w_6(\rho)$ shows how D7 probe brane will be bended by the  D3 brane (or $AdS_5\times S^5$).  This will code the behavior of  quark in the corresponding  field theory. 

    Note that, after adding the  D7 brane probe there are the light modes coming from strings with one end on the D3-branes and the other one on the D7-brane, which will give rise to quark hypermultiplet in the fundamental representation.  In this case,  if the D3-branes and the D7-brane overlap the hypermultiplet will be  massless as the distance between D7-brane and D3-branes is proportional to the hypermultiplet mass [4].   

   To proceed, we first know that the dynamics of the D7 brane probe is described by the combined Dirac-Born-Infeld and Chern-Simons actions,
$$S_{D7} = -\mu_7\int d^8 \xi e^{-\Phi}\sqrt{-det(P[G+ B]_{ab})} + \mu_7\int  P[C^{(4)}]\wedge B\wedge B,\eqno{(3.4)}$$
in which $\mu_7$ is the D7-brane tension and $P$ denotes the pullback of a bulk field to the world-volume of the brane. 

 To investigate the chiral dynamics we will let $w_5$  and $w_6$ to be function of $\rho$.  After substituting the background (3.1), (3.2) into (3.4) we find that the action for a static D7 embedding is given by
$$S_{D7} = -\mu_7\int d^8 \xi ~\rho^3 \sqrt{1+ \left({\partial w_5\over \partial \rho }\right)^2+ \left({\partial w_6\over \partial \rho }\right)^2},\eqno{(3.5)}$$
up to angular factors.   As the eq.(3.5) is just that without dipole field [2,3] and we thus conclude that non-commutative dipole field does not induce the chiral symmetry breaking in the supersymmetric theory. 

To investigate the meson spectrum we first let  $w_5$  and $w_6$ to be function of ($\rho, x,y,z) $ [4]. The associated Lagrangian is
$$ {\cal  L} = - {\rho^3} \left[1+ \left(\partial_ \rho \vec w \right)^2 +{\left(\partial_x  \vec w\right)^2+ \left(\partial_y \vec w\right)^2+ \left(\partial_z \vec w\right)^2\over (\rho^2+ \vec w^2)^2}\right], \eqno{(3.6)}$$
in which $\vec w = (w_5,w_6)$.   Now, to consider the quadratic fluctuations we can let $\vec w = \vec L +  \delta \vec w $, in which $|\vec L|$ is the distance between the D3- and D7-branes and $\delta \vec w $ is the small fluctuation.  However, the  Lagrangian (3.6) is just that without dipole field and meson spectrum is $ M_0^2 = 4L^2(n+1)(n+2)$, in which quantum number $n=0,1..$.  Thus we conclude that the non-commutative dipole field does not modify the meson spectrum in the supersymmetric theory.   Note that authors in [17] had shown that the non-commutativity of  Moyal product could  modify the meson spectrum in the supersymmetric theory.   
\subsection{Non-supersymmetric dipole field theory}
After the evaluations  the following supergravity solution are found 
\\
$$ds_{10}^2 = \left(\rho^2 +  \vec w^2\right)\left[-dt^2+ dx^2+ dy^2+{ dz^2\over 1+B^2\left(\rho^2 +  \vec w^2\right)}\right]+{1\over \rho^2 +  \vec w^2} \left[ d \vec w^2 + d\rho^2 \right.\hspace{4cm}$$
$$ + \rho^2 \left(d\theta^2+ \cos^2\theta d\phi^2 + \sin^2\theta d\chi^2\right)  - {B^2 \left[ \rho^2 \left(\cos^2\theta d\phi + \sin^2\theta d\chi\right) +   (w_5dw_6 -w_6dw_5)\right]^2\over 1+B^2\left(\rho^2 +  \vec w^2\right)}. \eqno{(3.7)}$$
$$e^{2\Phi}= {1 \over  1+B^2\left(\rho^2 +  \vec w^2\right)},~~~
 \sum_{n_i=\phi,\chi...} B_{zi}dn_i= - {B\left[ \rho^2 \left(\cos^2\theta d\phi + \sin^2\theta d\chi\right) + (w_5dw_6 - w_6dw_5)\right] \over 1+B^2\left(\rho^2 +  \vec w^2\right)},\hspace{4cm}\eqno{(3.8)}$$
in the large \textsf{N} limit.

First, we can follow the previous prescription to investigate the chiral dynamics.  It is found that the action for a static D7 embedding is still given by (3.5).  Thus, we conclude that non-commutative dipole field does not induce the chiral symmetry breaking in the non-supersymmetric theory, , in contrast to the general believing that the breaking supersymmetry in gravity duals will leads to chiral symmetry breaking [16]. 

Next, we can also follow the previous prescription to investigate the meson spectrum,  The associated Lagrangian is 
$$ {\cal  L} = - {\rho^3} \left[1+ \left(\partial_ \rho \vec w \right)^2 +{\left(\partial_x  \vec w\right)^2+ \left(\partial_y \vec w\right)^2+ \left(\partial_z \vec w\right)^2\over (\rho^2+ \vec w^2)^2}\right.\hspace{5cm}$$
$$\hspace{3cm}\left. + { B^4 \vec w^2 \left(\partial_z \vec w\right)^2\over (\rho^2+ \vec w^2)^2} -  { B^2 \left(1+ B^2 \vec w\right)  \over (\rho^2+ \vec w^2)^2\left(1+ B^2\rho^2\right)} {\left(w_6\partial_z w_5 -w_5\partial_z w_6  \right)^2}\right].  \eqno{(3.9)}$$
Now,  consider $(w_5,w_6) = (0+ \delta w_5, L)$  with fluctuation $\delta w_5 = G(\rho,z) = g(\rho) e^{ikz}$ we find that the associated equation for the field $g(\rho)$ becomes
$$(\rho^2+L^2)^2 \partial^2_\rho g(\rho) +{3(\rho^2+L^2)^2\over \rho}\partial_\rho g(\rho) - \left(M^2 +\gamma ^2B^4 L^2 M^2 \rho^2 \right)g(\rho)=0,\eqno{(3.10)}$$
to the leading order of small value of dipole field $B$, which means that it is compared to the radius of the undeformed $S^5$ radius $R_{S}$.  The meson spectrum is  $M^2 \equiv -k^2$.  Note that the dipole field will deform $S^5$ and, for simplicity, we  merely consider the mode the $S$-orbital states.

Now, following the method of [4] we could solve above equation by making the substituting
$$ g(\rho) = (\rho^2+L^2)^{-\alpha} P(\rho),\eqno{(3.11)}$$
and using the variable $y=-\rho^2/L^2$ to express Eq.(3.4)  
$$(1-y)^2 yP''(y) + 2(1-y) \left(1-(1-\alpha)y\right) P'(y) +\left[\left(2\alpha- {M^2\over4}\right)-\left(\alpha-\alpha^2 +{M^2B^4\over4}\right)y\right]P(y)=0.\eqno{(3.12)}$$
Above equation has solution which could be expressed as the hypergeometric function [4] and we find that  
$$ \alpha ={(n+1)\left(1-(n+2) B^4L^2\right)},~~~~ P(\rho) =F[-n-1,-n:2, -\rho^2/L^2],\eqno{(3.13)}$$
$$~~~ M^2 = M_0^2  + \Delta M^2, ~~~~ with~~~\Delta M^2= - 4  B^4L^2(n+1)(n+2).\eqno{(3.14)}$$
Eq. (3.8) tells us that the dipole field will reduce the meson energy  in the non-supersymmetric theory and that the correction term is an increasing function of quantum number $n$. 

Note that for the theory in the $AdS$ background the symmetry property in  $AdS$ lead us to have the same meson spectrum whether the fluctuation $\delta w$ is chosen as $\sim e^{ikt}$, $\sim e^{ikx}$, $\sim e^{iky}$ or $\sim e^{ikz}$. However, in our case as the $AdS$ background has been deformed by the $B$ field the relevant symmetry was broken, as the line element appears factor ${1\over 1+B^2\left(\rho^2 +  \vec w^2\right)}$ in $dz^2$.  In choosing  $\delta w \sim e^{ikt}$, $\sim e^{ikx}$ or $\sim e^{iky}$ we  have the spectrum as that without dipole field.  Only choosing  $\delta w \sim e^{ikz}$ could it give the spectrum that shows dipole deformation effect.

\section{Quark and anti-quark potential}
We will evaluate the potential energy for a static quark-antiquark pair in the non-commutative dipole field theory form dual gravity approach as that in [20].  Following [4] we will consider a static configuration consisting of a string stretched in the $z$ direction with both ends attached to the D7-brane probe placed
at $\rho = \rho_{D7}$. The relevant part of the metric from (3.1) in supersymmetric theory or form (3.7) in non-supersymmetric theory is 
$$ds^2= \rho^2\left[-dt^2+ {dz^2\over 1+B^2 \rho^2}\right] + {d\rho^2\over \rho^2}.\eqno{(4.1)}$$
Using $z$ as worldvolume coordinate and considering an ansatz of the form $\rho = \rho(z)$, the Nambu-Goto action reads
$${\cal L} = -\sqrt{{\rho^4\over 1+B^2 \rho^2}+ \left({d\rho\over dz}\right)^2}.\eqno{(4.2)}$$
As ${\cal L}$ does not depend explicitly on $z$, the quantity ${\rho'}{\partial{\cal L}\over \partial\rho'}-{\cal L}$ is a constant and we find that 
$${{\rho^4\over 1+B^2\rho^2}\over \sqrt{{\rho^4\over 1+B^2 \rho^2}+ \rho'^2}} = {\rho_0^2\over \sqrt{1+B^2\rho_0^2}},~~~~\Rightarrow~~~\left({d\rho\over dz} \right)^2 = {\rho^4(\rho^2-\rho_0^2)(\rho^2+\rho_0^2+ B^2\rho_0^2\rho^2)\over \rho_0^4(1+B^2\rho^2)^2}\eqno{(4.3)}$$
in which $\rho_0$  is a constant of motion. 

 Using above relation the quark-antiquark separation $L$ could be calculated by 
$$ L = {2}\int dz= {2}\int _{\rho_0}^{\rho_{D7}/\rho_0}\left({d\rho\over dz}\right)^{-1} d\rho={2\over \rho_0}\int _1^{\rho_{D7}/\rho_0}{1+B^2 \rho_0^2 y^2\over y^2 \sqrt{(y^2-1)(y^2+1+B^2 \rho_0^2 y^2)}}~ dy.\eqno{(4.4)}$$
In a similar way, the energy for this static configuration becomes
$$ H =\int \sqrt{{\rho^4\over 1+B^2 \rho^2}+ \left({d\rho\over dz}\right)^2}dz=\int _{\rho_0}^{\rho_{D7}/\rho_0}\sqrt{{\rho^4\over 1+B^2 \rho^2}+ \left({d\rho\over dz}\right)^2}\left({d\rho\over dz}\right)^{-1} d\rho
$$
$$ ={\rho_0}\sqrt{1+B^2 \rho_0^2}\int _1^{\rho_{D7}/\rho_0} {y^2\over  \sqrt{(y^2-1)(y^2+1+B^2 \rho_0^2 y^2)}}~ dy.\hspace{2.5cm}\eqno{(4.5)}$$
From above two relations we could find the quark-antiquark potential as a function of the distance between the quark and the antiquark. The results are in figure 1.  
\\
 
\hfil\scalebox{1}{\includegraphics{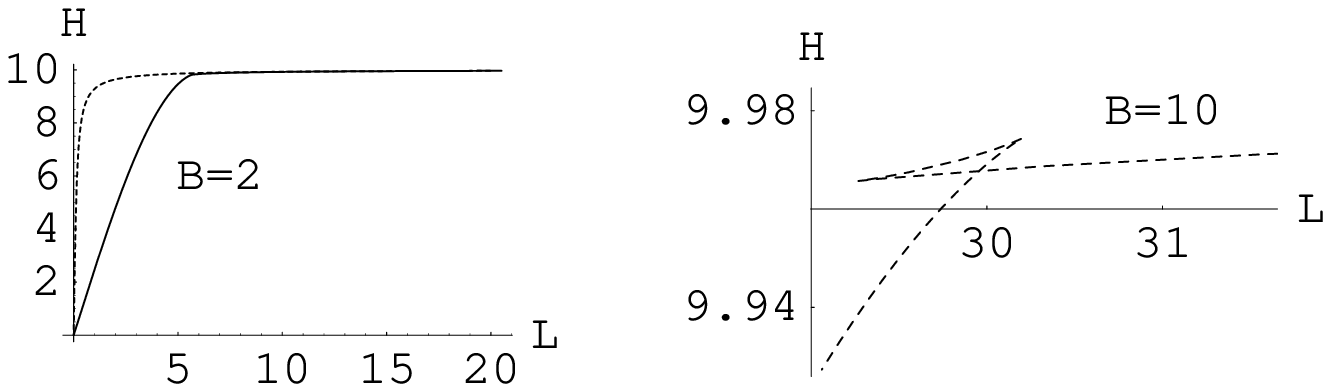}}\hfil\\
{\it Figure 1:  Potential of quark anti-quark pair.  Dashed line in left figure is that without dipole field.  When the dipole field is large then there will show a transition behavior at a critical distance $L_c$ as shown in right figure. This property also shows in  the Moyal non-commutative theory [17].} 
\\
\\
The figure 1 shows that the potential is linear for small quark anti-quark
separation, while for large separation the energy becomes constant and equal to $2m_q (=\rho_{D7})$.  The potential is a decreasing function of dipole field $B$.  The property could be explicitly seen in the small and long separation limits by the following analysis.

At the small separation limit $L \rightarrow 0$, i.e.  when $\rho_{0}\rightarrow\rho_{D7}$, then we can define $\rho_{D7}/\rho_{0}= 1+\epsilon$, and, at leading order, it is easily to see that
$$ L \rightarrow {1\over \rho_{D7}}{1+B^2 \rho_{D7}^2\over \sqrt{1+{1\over 2}B^2 \rho_{D7}^2}}\sqrt{{\rho_0\over \rho_{D7}}}.\eqno{(4.6)}$$
$$ H \rightarrow {\rho_{D7}\over 2}{\sqrt{1+B^2 \rho_{D7}^2}\over \sqrt{1+{1\over 2}B^2 \rho_{D7}^2}}\sqrt{{\rho_0\over \rho_{D7}}}.\eqno{(4.7)}$$
Thus we find the linear potential relation 
$$H(L)=\tau_{eff} L, ~~~~~~~with~~~~~\tau_{eff} ={\rho_{D7}\over 2} {1\over \sqrt{1+B^2 \rho_{D7}^2}},\eqno{(4.8)}$$
in which the effective tension  $\tau_{eff}$ is a decreasing function of the dipole field strength.

At long separation limit $L \rightarrow \infty$, i.e.  when $\rho_{0}\rightarrow 0$, the integration in (4.4) could be approximated as  
$$L = {2\over \rho_0}\int _1^{\rho_{D7}/\rho_0}{1+B^2 \rho_0^2 y^2\over y^2 \sqrt{(y^2-1)(y^2+1)\left(1+{B^2 \rho^2_0 y^2\over y^2+1}\right)}}~ dy\hspace{6cm}$$
$$\approx {2\over \rho_0}\int _1^{\rho_{D7}/\rho_0}\left[{1\over y^2 \sqrt{y^4-1}}-{B^2 \rho_0^2\over 2 (y^2+1) \sqrt{y^4-1}} +{B^2 \rho_0^2\over \sqrt{y^4-1}} \right]~ dy\hspace{3.5cm}$$
$$\approx {2\over \rho_0}\left[\int _1^{\rho_{D7}/\rho_0}{1\over y^2 \sqrt{y^4-1}}~dy-\int _1^\infty {B^2 \rho_0^2\over 2 (y^2+1) \sqrt{y^4-1}}~ dy +\int _1^\infty{B^2 \rho_0^2\over \sqrt{y^4-1}} ~ dy\right],\eqno{(4.9)}$$
in the case of  small dipole field. The first integral could be furthermore approximated by 
$$\int _1^{\rho_{D7}/\rho_0}{1\over y^2 \sqrt{y^4-1}}~dy =\int _1^{\infty}{1\over y^2 \sqrt{y^4-1}}~dy - \int _{\rho_{D7}/\rho_0}^\infty{1\over y^2 \sqrt{y^4-1}}~dy\approx C_1 - {\rho_{0}^3\over 3 \rho_{D7}^3},\eqno{(4.10)}$$
in which $C_1\approx 0.59907$. The values of second  and third integrals in (4.9) could be found and denoted as $C_2 \approx 0.3559$, $C_3 \approx 1.3110$.  We thus find that 
$$ L \approx {2 C_1\over \rho_0} +(2C_3-C_2)B^2\rho_0 -{2\over 3}{\rho_0^2\over \rho_{D7}^3}\eqno{(4.11)}$$
In a similar way we  find that 
$$H ={\rho_0}\sqrt{1+B^2 \rho_0^2} \int _1^{\rho_{D7}/\rho_0}{y^2\over  \sqrt{(y^2-1)(y^2+1)\left(1+{B^2 \rho^2_0 y^2\over y^2+1}\right)}}~ dy\hspace{6cm}$$
$$\approx {\rho_0}\sqrt{1+B^2 \rho_0^2} \int _1^{\rho_{D7}/\rho_0}\left[{y^2\over \sqrt{y^4-1}}-{B^2 \rho_0^2 y^4 \over 2 (y^2+1) \sqrt{y^4-1}}\right]~ dy\hspace{5.6cm}$$
$$= {\rho_0}\sqrt{1+B^2 \rho_0^2}\left[\int _1^{\rho_{D7}/\rho_0}~1~dy+\int _1^{\rho_{D7}/\rho_0}\left(-1 +{y^2\over  \sqrt{y^4-1}}\right)~dy-\int _1^{\rho_{D7}/\rho_0} {B^2 \rho_0^2\over 2 (y^2+1) \sqrt{y^4-1}}~ dy \right] $$
$$= {\rho_0}\sqrt{1+B^2 \rho_0^2}\left[{\rho_{D7}\over\rho_0} + \left[-1 +\int _1^{\infty}\left(-1 +{y^2\over  \sqrt{y^4-1}}\right)~dy\right] - \left[\int _{\rho_{D7}/\rho_0}^{\infty}\left(-1 +{y^2\over  \sqrt{y^4-1}}\right)~dy \right] \right.$$
$$\left.-\int_1^{\rho_{D7}/\rho_0} {B^2 \rho_0^2\over 2 (y^2+1) \sqrt{y^4-1}}~ dy \right]  \approx \rho_{D7} -  C_1 \rho_0 -{B^2\rho_0^2 \rho_{D7}\over 4}.\hspace{3cm}\eqno{(4.12)}$$
Solve (4.11) and (4.12) we find that the quark and anti-quark potential at long distance is 
$$H(L)= 2m_q - {2C_1 \over L} - {2 B^2C_1 m_q\over L^2} -{4 B^2C_1^3(2C_3-C_2)\over L^3},\eqno{(4.13)}$$
in which $\rho_{D7} = 2m_q$ and $m_q$ is the quark mass.  Eq.(4.8) and (4.12) tell us that the dipole field has an effect to produce attractive force between the quark and anti-quark.   Note that the energy corrected  by non-commutative dipole field at long distance  is shown in $L^{-2}$ while that by Moyal product is shown in $L^{-4}$ [17].

\section{Conclusion}
In conclusion, we have investigated the problem of chiral symmetry breaking and the meson spectroscopy in the non-commutative dipole field form the  gauge/gravity dual.   We first following the method in [13] to derive the dual supergravity background in the proper coordinate.  We find that the dipole field does not induce the chiral symmetry breaking in the supersymmetric and non-supersymmetric theory, in contrast to the general believing that the breaking supersymmetry in gravity duals will lead to chiral symmetry breaking [16]. We also find that the dipole field does not modify the meson spectrum in the supersymmetric theory, in contrast to that studied in [17] in which the meson spectrum could be modified by the space non-commutativity of Moyal product.   The dipole field will reduce the meson energy in the non-supersymmetric dipole field theory. Finally, we also evaluate the potential energy for a static quark-antiquark pair and see that the dipole field has an effect to produce attractive force between the quark and anti-quark.  Finally, the property of  chiral dynamics and meson on the other dipole field deformed background  [21] is of interesting and remains to be  studied.
\\
\\
\\
\begin{center} {\bf REFERENCES}\end{center}
\begin{enumerate}
\item J.~M.~Maldacena, ``The large N limit of superconformal field theories and supergravity,'' Adv. Theor. Math.  Phys. 2 (1998) 231 [hep-th/9711200];
 S. S. Gubser, I. R. Klebanov and A.~M.~Polyakov, ``Gauge theory correlators from non-critical string theory,'' Phys. Lett. B428 (1998) 105 [hep-th/9802109]; E.~Witten, ``Anti-de Sitter space and holography,'' Adv. Theor. Math. Phys. 2 (1998) 253 [hep-th/9802150].
\item  A. Karch and E. Katz, ``Adding avor to AdS/CFT", JHEP 06 (2002) 043,
[hep-th/0205236];  A. Karch, E. Katz, and N. Weiner, ``Hadron masses and screening from AdS Wilson loops", Phys. Rev. Lett. 90 (2003) 091601 [hep-th/0211107].
\item  J. Babington, J. Erdmenger, N. J. Evans, Z. Guralnik, and I. Kirsch, ``Chiral symmetry breaking and pions in non-supersymmetric gauge / gravity duals", Phys. Rev. D69 (2004) 066007 [hep-th/0306018]; N. J. Evans and J. P. Shock, ``Chiral dynamics from AdS space", Phys. Rev. D70 (2004) 046002 [hep-th/0403279].
\item M. Kruczenski, D. Mateos, R. C. Myers, and D. J. Winters, ``Meson spectroscopy in AdS/CFT with  flavour", JHEP 07 (2003) 049 [hep-th/0304032].
\item   N. R. Constable and R. C. Myers, ``Exotic scalar states in the AdS/CFT correspondence,"  JHEP 9911 (1999) 020 [hep-th/9905081]
\item  M. Kruczenski, D. Mateos, R. C. Myers, and D. J. Winters, ``Towards a holographic dual of large-N(c) QCD", JHEP 05 (2004) 041 [hep-th/0311270].
\item K. Ghoroku, M. Ishihara, and A. Nakamura, ``Flavor quarks in AdS(4) and gauge/gravity correspondence", Phys. Rev. D75 (2007) 046005 [hep-th/0612244].
\item S. Penati, M. Pirrone, C. A. Ratti, ``Mesons in marginally deformed AdS/CFT", arXiv:0710.4292v1 [hep-th].
\item  O. Lunin and J.M. Maldacena, ``Deforming Field Theories with $U(1)\times U(1)$ global symmetry and their gravity duals,"  JHEP 0505 (2005) 033 [hep-th/0502086]; S.A. Frolov, ``Lax Pair for Strings in Lunin-Maldacena Background," JHEP 0505 (2005) 069 [hep-th/0503201].
\item V. G. Filev, C. V. Johnson, R. C. Rashkov, and K. S. Viswanathan, ``Flavoured large N gauge theory in an external magnetic field", [hep-th/0701001].
\item  J. Erdmenger, R. Meyer, J. P. Shock, `` AdS/CFT with Flavour in Electric and Magnetic Kalb-Ramond Fields", JHEP 0712 (2007) 091, arXiv:0709.1551 [hep-th].
\item   J. Erdmenger, N. Evans, I. Kirsch, and E. Threlfall, `` Mesons in Gauge/Gravity Duals - A Review,''  arXiv:0711.4467 [hep/th].
\item A. Bergman and O. J. Ganor,``Dipoles, Twists and Noncommutative Gauge Theory," JHEP 0010 (2000) 018 [hep-th/0008030]; A. Bergman, K. Dasgupta, O. J. Ganor, J. L. Karczmarek, and G. Rajesh,``Nonlocal Field Theories and their Gravity Duals," Phys.Rev. D65 (2002) 066005 [hep-th/0103090]; M. Alishahiha and H. Yavartanoo,``Supergravity Description of the Large N Noncommutative Dipole Field Theories," JHEP 0204 (2002) 031 [hep-th/0202131] . 
\item K. Dasgupta and M. M. Sheikh-Jabbari, ``Noncommutative Dipole Field Theories," JHEP 0202 (2002) 002 [hep-th/0112064].
\item Wung-Hong Huang, ``Wilson-t'Hooft Loops in Finite-Temperature Non-commutative Dipole Field Theory from Dual Supergravity,''   Phys.Rev. D76 (2007) 106005, arXiv: 0706.3663 [hep-th]; ``Thermal Giant Graviton with Non-commutative Dipole Field", JHEP 0711 (2007) 015, arXiv: 0709.0320 [hep-th]
\item  N. Evans, J. Shock, and T. Waterson, ``D7 Brane Embeddings and Chiral Symmetry Breaking,'' JHEP 0503 (2005) 005 [hep-th/0502091].
\item  D. Arean, A. Paredes, and A. V. Ramallo, ``Adding flavor to the gravity dual of non-commutative gauge theories", JHEP 08 (2005) 017 [hep-th/0505181].
\item N. Sadooghi and M. Soroush, ``Noncommutative  Dipole QED", Int. J. Mod. Phys. A18 (2003) 97 [hep-th/0206009]. 
\item G.T. Horowitz and A.~Strominger, ``Black strings and P-branes,'' Nucl. Phys. B  360 (1991) 197.
\item J. M. Maldacena, Wilson loops in large N field theories??, Phys. Rev. Lett. 80 (1998) 4859 [hep-th/9803002]; S. J. Rey and J. Yee, Macroscopic strings as heavy quarks in large N theory and anti-de Sitter supergravity??, Eur. Phys. J. C22 (2001) 379 [hep-th/9803001].
\item U. Gursoy and C. Nunez   ,  ``Dipole Deformations of N=1 SYM and Supergravity backgrounds with U(1) X U(1) global symmetry ,'' Nucl.Phys. B725 (2005) 45-92  [hep-th/0505100 ]; N.P. Bobev, H. Dimov, R.C. Rashkov ,  ``Semiclassical Strings, Dipole Deformations of N=1 SYM and Decoupling of KK Modes,'' JHEP 0602 (2006) 064 [hep-th/0511216].

\end{enumerate}
\end{document}